\documentclass[a4,11pt]{article}

\usepackage[dvips]{graphics}
\usepackage{graphicx}
\usepackage{overcite}
\usepackage{amssymb}
\usepackage{amsmath}

\def\ds{\displaystyle}

\def\ie{{\rm i.e.}}
\def\I{{\rm i}} 
\def\e{{\rm e}} 
\def\TrB{{\rm Tr}_{\rm B}}
\def\deltaFunction{{\rm \delta}}
\def\H{{\cal H}} 
\def\HS{\H_{\rm S}}
\def\HB{\H_{\rm B}}
\def\HSB{\H_{\rm SB}}
\def\Hp#1{\H_{\rm p}(#1)}
\def\HWt{\tilde\H}
\def\HSBt{\HWt_{\rm SB}}
\def\HSBH#1{\hat\H_{\rm SB}(#1)}
\def\lSB#1{L_1 (#1)}

\def\LSBt#1{\tilde L_{\rm SB}(#1)}
\def\k{k}          
\def\kB{k_{\rm B}} 
\def\om{\omega}
\def\ok{\om_\k}
\def\oc{\om_{\rm c}}
\def\tc{t_{\rm c}}
\def\Dt{\Delta t}
\def\dt{\delta t}
\def\np{N_{\rm p}}
\def\ihb{\frac{\I}{\hbar}}
\def\s{\sigma} 

\def\sz{\s_{\it z}}
\def\rW#1{\rho(#1)}
\def\rB{\rho_{\rm B}}
\def\rWt#1{{\tilde\rho}(#1)}
\def\rSt#1{{\tilde\rho}_{\rm S}(#1)}
\def\rBt#1{{\tilde\rho}_{\rm B}(#1)}
\def\rt#1{{\tilde\rho}_{#1}}
\def\nB{n_{\rm B}(\om)}
\def\nBk{n_{\rm B}(\ok)}
\def\roundy#1{\frac{\partial}{\partial#1}}
\def\Ket#1{\lvert#1\rangle}     
\def\abs#1{\lvert#1\lvert}      

\title
{\Large Pulse Control of Decoherence in a Qubit Coupled with a Quantum Environment}

\author
{
Takahiro {\sc Murakami}\thanks{E-mail address: m1279028@hiroshima-u.ac.jp}
and Yositake {\sc Takane}
\\
{\small Department of Quantum Matter, ADSM, Hiroshima University,}
\\
{\small 1-3-1 Kagamiyama, Higashi-Hiroshima, Hiroshima 739-8530, Japan}
}

\begin{document}

\maketitle
\begin{abstract}
We study the time evolution of a qubit linearly coupled
with a quantum environment under a sequence of short $\pi$ pulses.
Our attention is focused on the case
where qubit-environment interactions induce the decoherence
with population decay.
We assume that the environment consists of a set of bosonic excitations.
The time evolution of the reduced density matrix for the qubit is
calculated in the presence of periodic short $\pi$ pulses.
We confirm that the decoherence is suppressed
if the pulse interval is shorter than the correlation time
for qubit-environment interactions.
\end{abstract}

\section{Introduction}
To employ a two-level quantum system as a qubit for quantum computations,
phase coherence must be well preserved in it for a long time.
However, since any quantum systems interact with their environment,
decoherence inevitably arises more or less.
Thus, several schemes to suppress decoherence have been proposed
\cite{Various1,Various2,Ban,ViolaLloyd}.
Among them, the decoherence suppression scheme
which uses a sequence of short $\pi$ pulses
\cite{Ban,ViolaLloyd,DuanGuo,VitaliTombesi,UchiyamaAihara}
is a promising candidate for actual applications
since it is relatively easy to implement in experimental situations
and can be readily applied to multi-qubit systems.
We call this scheme the pulse control of decoherence.
The pulse control has been proposed by Ban \cite{Ban},
and by Viola and Lloyd \cite{ViolaLloyd}.
They considered a qubit under a sequence of short $\pi$ pulses
assuming that qubit-environment interactions induce the decoherence
without population decay (\ie, dephasing).
They calculated the time evolution of the qubit,
and showed that the dephasing can be suppressed
by applying periodic $\pi$ pulses
if the pulse interval is shorter than the correlation time
for qubit-environment interactions.
It has been proposed that we can even suppress the decoherence
with population decay by applying $\pi$ pulses \cite{DuanGuo}.
Let us consider a qubit linearly interacting with a boson environment.
The Hamiltonian is $\H_0 = \HS +\HB +\HSB$ with
\begin{eqnarray}
\HS &= &\hbar\om_0\frac{\sz}{2} \otimes 1_{\rm B}
\label{HS},
\\
\HB &= &1_{\rm S} \otimes \sum_k \hbar\ok b_k^\dagger b_k
\label{HB}.
\end{eqnarray}
Here, $\HS$ and $\HB$ describe the qubit and the environment, respectively,
and $\HSB$ describes their mutual interactions.
We have used the pseudo-spin representation in expressing the qubit.
Ban, and Viola and Lloyd, considered the case where $\HSB$ is given by
\begin{equation}
\HSB =\sum_k
        \hbar \left(
               g_k^* \sz \otimes b_k^\dagger
               +g_k  \sz \otimes b_k
              \right)
\label{HSBz},
\end{equation}
which induces the dephasing.
They proposed that the dephasing can be suppressed
by applying periodic $\pi$ pulses around the $x$-axis in the pseudo-spin space.
Let $\rW{t}$ be the density matrix for the whole system
consisting of the qubit and the environment.
We can simply integrate
the Liouville-von Neumann equation for $\rW{t}$ under the $\pi$ pulses
since $\left[ \HS,\HSB \right] = 0$,
and explicitly examine the effectiveness of the pulse control.
In contrast, if
\begin{equation}
\HSB =\sum_k
        \hbar \left(
               g_k^* \s_- \otimes b_k^\dagger
               +g_k  \s_+ \otimes b_k
              \right)
\label{HSB},
\end{equation}
the decoherence with population decay arises.
It has been proposed that the decoherence with population decay
can be suppressed by a sequence of $\pi$ pulses around the $z$-axis
\cite{DuanGuo}.
However, its effectiveness has not been quantitatively examined so far.

Generally, a qubit experiences
not only the dephasing but also the decoherence with population decay.
Thus, it is desirable to examine the effectiveness of the pulse control
in more general situations beyond the case
treated in refs.
\citenum{Ban} and \citenum{ViolaLloyd}.
In this paper, we present a formulation
to describe the time evolution of the qubit coupled with a quantum environment
under a sequence of $\pi$ pulses.
Our formula is based
on the time-convolutionless (TCL) projection operator approach,
and is applicable to the case
where the decoherence with population decay is present.
An extension to more general cases
where both the dephasing and the decoherence with population decay
play a role
is straightforward.

\section{Formulation}
Let us consider the case
where the qubit-environment interaction is given by eq. (\ref{HSB})
and periodic $\pi$ pulses with an interval $\Dt$ are applied.
Let $\dt$ be the duration of each pulse.
The influence of the $\pi$ pulses is described by
$\Hp{t}=V(t) \sz \otimes 1_{\rm B}$, where
\begin{equation}
V(t)=\left\{ \hspace{-1mm}
      \begin{array}{cl}\ds
        0 & \mbox{for } m(\Dt+\dt)      \le t <m(\Dt+\dt)+\Dt ,
        \\ \ds
        \frac{\pi \hbar}{2\dt}
          & \mbox{for } m(\Dt+\dt)+\Dt  \le t <(m+1)(\Dt+\dt)
      \end{array}
     \right.
\label{Vt}
\end{equation}
with $m=0,1,2,\cdots$.
Thus, the total Hamiltonian is $\H(t) \equiv \H_0 +\Hp t$.
We decompose $\H(t)$ as $\H(t)=\H_1(t)+\HSB$,
where $\H_1(t)=\HS+\HB+\Hp{t}$.
The whole system is described by the density matrix $\rW{t}$,
which obeys the Liouville-von Neumann equation
\begin{equation}
\roundy{t}\rW{t} = L(t)\rW{t}
\label{LiouvilleEq.}
\end{equation}
with 
\begin{equation}
L(t) \,\cdot\, =-\ihb\left[ \H(t), \,\cdot\, \right]
\label{LiouvilleOperator}.
\end{equation}
It is convenient to introduce $\rWt{t}$ which is defined as
\begin{equation}
\rWt{t}=T_{\rm\!a}\e^{-\int_0^t\!d\tau \lSB{\tau}} \rW{t},
\end{equation}
where $T_{\rm\!a}$ is the anti-time-ordering operator
and $\lSB t \cdot = -\ihb [\H_1(t),\cdot]$.
From eq. (\ref{LiouvilleEq.}), we obtain
\begin{equation}
\roundy{t}\rWt{t} = \LSBt{t}\rWt{t}
\label{LiouvilleEq.t},
\end{equation}
where $\LSBt t \cdot = -\ihb [\HSBt(t),\cdot]$.
Here, $\HSBt(t)$ is given by
\begin{equation}
\HSBt(t)=T_{\rm\!a}\e^{\ihb\int_0^t\!d\tau \H_1(\tau)}
          \HSB T\e^{-\ihb\int_0^t\!d\tau \H_1(\tau)}
\label{HSBt},
\end{equation}
where $T$ is the time-ordering operator.
We derive an explicit expression of $\HSBt(t)$.
Noting that $\H_1(t)=\HS+\HB+\Hp{t}$ and $[\HS+\HB,\Hp{t}]=0$,
we obtain
\begin{equation}
\HSBt(t)=T_{\rm\!a}\e^{\ihb\int_0^t\!d\tau \Hp\tau}
          \HSBH{t}T\e^{-\ihb\int_0^t\!d\tau \Hp\tau}
\label{HSBt2},
\end{equation}
where
\begin{eqnarray}
\HSBH{t}
 &= &\e^{\ihb(\HS+\HB)t} \HSB \e^{-\ihb(\HS+\HB)t}
\nonumber
\\
 &=
 &\sum_k
    \hbar \left(
           g_k^* \s_- \otimes b_k^\dagger \e^{\I(\ok-\om_0)t}
           +g_k  \s_+ \otimes b_k        \e^{-\I(\ok-\om_0)t}
          \right)
\label{HSBH}.
\end{eqnarray}
We here take the limit of $\dt\rightarrow0$.
This approximation is justified as long as $\dt$ is much shorter
than the typical time scales for the decoherence.
In the interval between the $\np$th pulse and the $(\np+1)$th pulse
(\ie, $\np\Dt<t<(\np+1)\Dt$ ),
we find that
\begin{equation}
\ihb\int_0^t\!d\tau \Hp{\tau} =\I\frac{\pi}{2}\np \sz\otimes1_{\rm B}.
\end{equation}
We thus obtain
\begin{eqnarray}
\HSBt(t) &= &\e^{\I\frac{\pi}{2}\np \sz\otimes1_{\rm B}}
              \HSBH t \e^{-\I\frac{\pi}{2}\np \sz\otimes1_{\rm B}}
\nonumber
\\
         &= &(-1)^{\np} \HSBH t
\label{npHSBt},
\end{eqnarray}
which results in
\begin{equation}
\LSBt t \,\cdot\, = -(-1)^{\np}\ihb[\HSBH t, \,\cdot\, ]
\label{npLSBt}.
\end{equation}
Note that each $\pi$ pulse changes the sign of $\LSBt t$.
Due to the periodic sign changes,
the qubit-environment interaction is effectively reduced.
This is the reason why the decoherence is suppressed by the $\pi$ pulses.

Our interest is focused on the time evolution of the qubit,
which is described by the reduced density matrix
defined by
\begin{equation}
\rSt t =\TrB \{\rWt t \}
=\begin{pmatrix}
  \rt{11}(t) & \rt{10}(t) \cr
  \rt{01}(t) & \rt{00}(t)
 \end{pmatrix} ,
\end{equation}
where
$\rt{11}$ ($\rt{00}$) represents the population in the upper (lower) state
and the off-diagonal terms characterize the coherence of the qubit.
In deriving the equation of motion for $\rSt{t}$,
we assume that the qubit and the environment are uncorrelated at $t=0$, \ie,
\begin{equation}
\rWt{0}=\rSt{0} \otimes \rBt{0}
\label{initial},
\end{equation}
and the environment is initially in thermal equilibrium
at temperature $T$, \ie,
\begin{equation}
\rBt{0}=\rB
       =\prod_k \left( 1-\e^{-\beta\hbar\ok} \right)
          \e^{-\beta\hbar\ok b_k^\dagger b_k }
\label{Bath}
\end{equation}
with $\beta=1/k_{\rm B}T$ ($k_{\rm B}$: the Boltzmann constant).
We introduce the projection operator $P$
which is defined as $P\cdot=\TrB \{\cdot\}\otimes\rB$.
We observe that $P\rWt{t}=\rSt{t}\otimes\rB$.
Adapting the TCL formalism \cite{ShibataArimitsu}
with $P$ to eq. (\ref{LiouvilleEq.t}),
we can derive the equation of motion for $\rSt{t}$,
\begin{equation}
\roundy{t}\rSt{t}
=\TrB \sum_{j=1}^\infty K_j(t) P\rWt{t}
\label{ExpanssionFormula},
\end{equation}
where
\begin{equation}
K_j(t)=\int_0^{t}dt_1
        \int_0^{t_1}dt_2
         \cdots \int_0^{t_{j-2}}dt_{j-1}
          \langle
           \LSBt{t}\LSBt{t_1}\cdots \LSBt{t_{j-1}}
          \rangle_{\rm o.c.}
\label{K}.
\end{equation}
Here, $\langle\cdots\rangle_{\rm o.c.}$ denotes an ordered cumulant
\cite{ShibataArimitsu}
\begin{eqnarray}
\langle \LSBt{t}\cdots \LSBt{t_{j-1}} \rangle_{\rm o.c.}
= \sum
   (-1)^{M-1}
    \prod
     P\LSBt{t}\cdots\LSBt{t_{q1}}
      P\LSBt{t_{q2}}\cdots \LSBt{t_{q3}}P\LSBt{t_{q4}}\cdots .
\label{OrderedCumulant}
\end{eqnarray}
The sum in eq. (\ref{OrderedCumulant}) should be taken
over all possible divisions by $P$ keeping the chronological order
specified by $t>\cdots>t_{q1}$, $t_{q2}>\cdots>t_{q3}$, $\cdots$ and so on,
and $M$ is the number of $P$ in each term.

Note that $K_j(t)P\rWt{t}$ with an odd $j$ vanishes in our case.
The qubit-environment interaction is not strong in actual situations,
so we are allowed to retain only a few lower-order terms
in the right-hand side of eq. (\ref{ExpanssionFormula}).
For simplicity, we here approximately neglect
the higher-order terms $K_j(t)$ with $j\ge4$.
The role of such higher-order terms will be discussed elsewhere.
We thus obtain
\begin{eqnarray}
\roundy{t}\rSt{t}
 &= &\TrB K_2(t) P\rWt{t}
\nonumber
\\
 &= &\TrB \left\{ \LSBt{t}\LSBt{t_1} \bigl\{\rSt{t}\otimes\rB\bigr\} \right\}
\label{SLiouvilleEq.t}.
\end{eqnarray}
Using eqs. (\ref{HSBH}) and (\ref{npLSBt}),
we obtain for $\np\Dt<t<(\np+1)\Dt$,
\begin{eqnarray}
\roundy{t}\rt{11}(t) &= &-\gamma_{11}(t)\rt{11}(t)+\eta_{11}(t)
\label{DifferentialRho11} ,
\\
\roundy{t}\rt{10}(t) &= &-\gamma_{10}(t)\rt{10}(t)
\label{DifferentialRho10} ,
\end{eqnarray}
where
\begin{eqnarray}
\gamma_{11}(t)
&= &\sum_k \abs{g_k}^2 \,\Bigl(2\nBk+1\Bigr)
      \int_{\!\np} dt_1 \,2\cos(\ok-\om_0)(t-t_1)
\label{gamma11} ,
\\
\gamma_{10}(t)
&= &\sum_k \abs{g_k}^2 \,\Bigl(2\nBk+1\Bigr)
      \int_{\!\np} dt_1 \e^{\I(\ok-\om_0)(t-t_1)}
\label{gamma10} ,
\\
\eta_{11}(t)
&= &\sum_k \abs{g_k}^2 \,\nBk
      \int_{\!\np} dt_1 \,2\cos(\ok-\om_0)(t-t_1)
\label{eta11} ,
\\
\int_{\!\np} dt_1 \cdots
&= &\int_{\np\Dt}^t dt_1 \cdots\,
    -\sum_{j=0}^{\np-1}(-1)^{j}
       \int_{(\np-1-j)\Dt}^{(\np-j)\Dt} dt_1\,\cdots
\label{IntNp} ,
\end{eqnarray}
and $\nB$ is the Bose-Einstein distribution function.
In deriving eq. (\ref{DifferentialRho11}),
we have used the relation of $\rt{11}(t)+\rt{00}(t)=1$.
Introducing the spectral function $I(\om)$ defined by
\begin{equation}
I(\om)=\sum_k \abs{g_k}^2 \deltaFunction(\om-\ok) ,
\end{equation}
we rewrite eq. (\ref{gamma11}) as
\begin{equation}
\gamma_{11}(t)
=\int_0^\infty d\om I(\om) \Bigl(2\nB+1\Bigr)
   \int_{\!\np} dt_1 2\cos(\om-\om_0)(t-t_1)
\label{gamma11Spectral}
\end{equation}
and eqs. (\ref{gamma10}) and (\ref{eta11}) are rewritten in similar manner.

\section{Results}
\begin{figure}[htbp]
\includegraphics*[scale=.41]{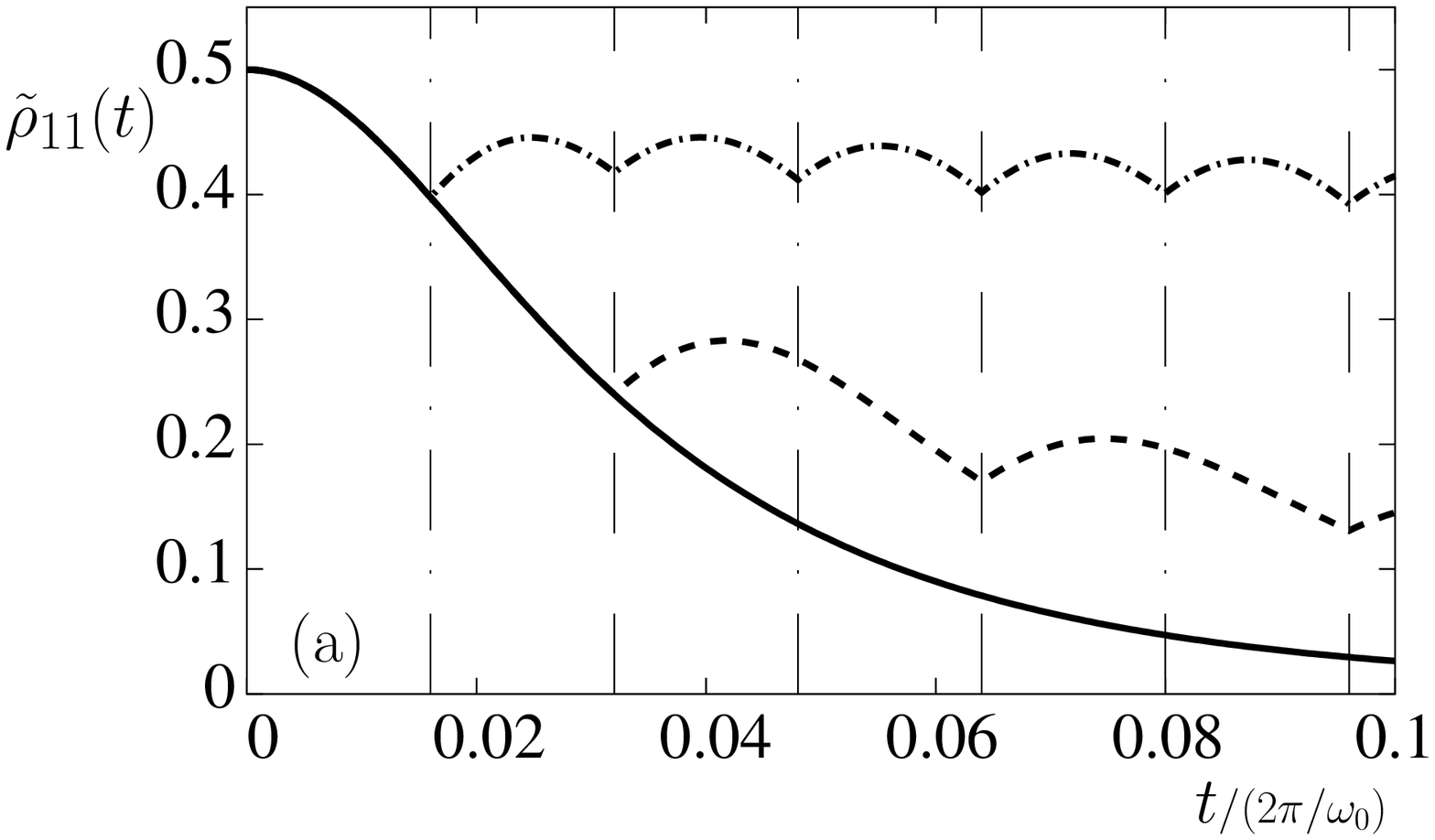}\hspace{10mm}
\includegraphics*[scale=.41]{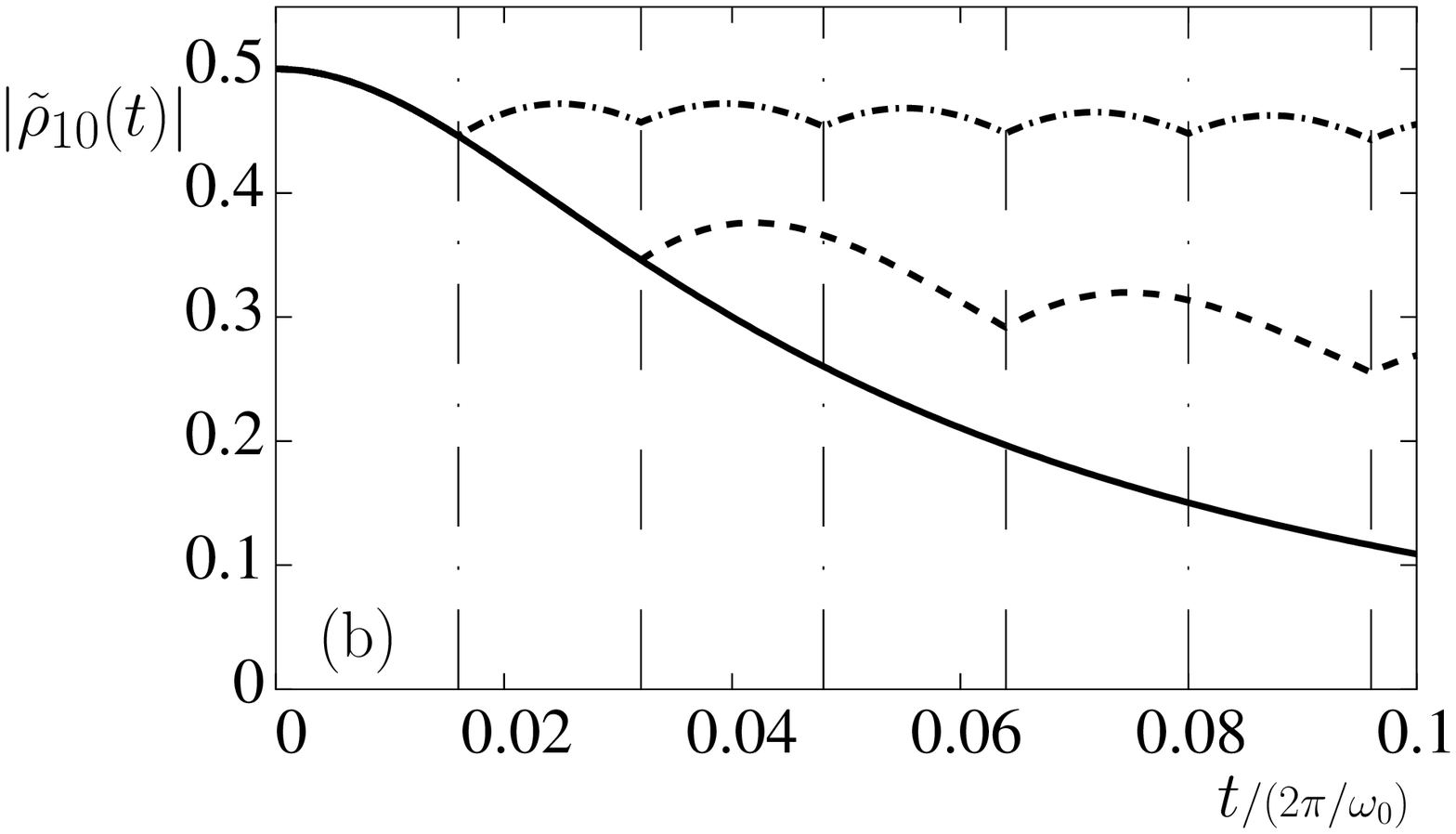}
\vspace{-3mm}
\caption
 {
  Time evolution of (a) $\rt{11}$ and (b) $\abs{\rt{10}}$
  at $\kB T=0.1\times\hbar\om_0$
  when the initial state is $(\Ket{0}+\Ket{1})/\sqrt{2}$.
  The dashed lines and dash-dotted lines
  correspond to the cases
  of $\Dt/(2\pi/\om_0)=0.016$ and $0.032$, respectively.
  The solid lines represent the case where $\pi$ pulses are absent.
 }
\label{ResultsGraph}
\end{figure}

On the basis of the resulting equations,
we study the time evolution of the reduced density matrix $\rSt{t}$
in the presence of periodic $\pi$ pulses with an interval $\Dt$.
We numerically calculate $\rt{11}(t)$ and $\rt{10}(t)$
assuming that the initial state is $(\Ket0+\Ket1)/\sqrt2,$ \ie,

\begin{equation}
\rSt0=\begin{pmatrix}
       0.5 & 0.5 \cr
       0.5 & 0.5
      \end{pmatrix},
\end{equation}
and that the spectral function is given by
\begin{equation}
I(\om)=\om \e^{-\om/\oc}
\label{Spectral},
\end{equation}
where $\oc$ is cut-off frequency.
The following parameters are employed:
$\oc/\om_0=5.0$ and $\kB T/(\hbar\om_0)=0.1$.
Then, the correlation time $t_{\rm c} \equiv 2\pi/\oc$
for qubit-environment interactions
is obtained as $\tc/(2\pi/\om_0)=0.2$.
If the $\pi$ pulses are absent,
$\rt{11}(t)$ approaches to
\begin{equation}
\rt{11}(\infty)=\frac
                 {\e^{-\hbar\om_0/(2\kB T)}}
                 {\e^{+\hbar\om_0/(2\kB T)} +\e^{-\hbar\om_0/(2\kB T)}}
               \sim O(10^{-5})
\end{equation}
with increasing $t$.
We treat the cases of $\Dt/(2\pi/\om_0)=0.016$ and $0.032$,
for which the condition $\Dt\ll \tc$ holds.
We display $\rt{11}(t)$ and $\abs{\rt{10}(t)}$
in Fig. \ref{ResultsGraph}(a) and \ref{ResultsGraph}(b), respectively.
Note that the decrease of $\rt{11}$ ($\abs{\rt{10}}$) represents
the population decay (decoherence).
From Fig. \ref{ResultsGraph}(a),
we observe that the population decay is suppressed by the $\pi$ pulses
and the suppression becomes notable with decreasing $\Dt$.
Figure \ref{ResultsGraph}(b) shows that
the coherence of the qubit is recovered by the $\pi$ pulses.

We apply the $\pi$ pulses around the $z$-axis,
so each pulse does not directly modify the population.
Note that reduction of the population decay is attributed
to the periodic sign changes of the qubit-environment interaction
caused by the $\pi$ pulses.
Here we mention to an another way to reduce the population decay.
The population decay can be suppressed without using $\pi$ pulses
if we simply raise $T$.
Indeed, since we have set $\rt{11}(0)=\rt{00}(0)=0.5$,
the population decay disappears in the extreme limit of $\kB T \gg \hbar\om_0$.
However, this inevitably accelerates the decoherence.

\section{Summary}
We have presented a formulation
to describe the time evolution of a qubit coupled with a quantum environment
under a sequence of short $\pi$ pulses.
We have applied it to the case
where the environment induces the decoherence with population decay,
and derived the equation of motion
for the reduced density matrix for the qubit.
By numerically solving the equation of motion,
we have obtained the time evolution of the reduced density matrix.
It is shown that, by applying periodic $\pi$ pulses,
we can suppress the decoherence with population decay
if the pulse interval is much shorter than the correlation time
for qubit-environment interactions.

\section*{Acknowledgement}
The present authors thank Prof. M. Yamanishi
for calling their attention to ref.
 \citenum{ViolaLloyd}.

\end{document}